\documentstyle[twoside,fleqn,espcrc2,epsfig]{article}
\newcommand{\AmS}{{\protect\the\textfont2
  A\kern-.1667em\lower.5ex\hbox{M}\kern-.125emS}}

\title{Disconnected Electromagnetic Form Factors}

\author{Walter Wilcox\address{Department of Physics, Baylor 
								University, Waco, TX 76798-7316}}
       
\begin{document}

\begin{abstract}
Preliminary results of a calculation of 
disconnected nucleon electromagnetic
factors factors on the lattice are presented. 
The implementation of
the numerical subtraction scheme is outlined. A comparison of
results for electric and magnetic disconnected form factors
on two lattice sizes with those of the Kentucky group
is presented. Unlike previous results, the results 
found in this calculation
are consistent with zero in these sectors.
 
\end{abstract}

\maketitle

\section{Introduction}

Fig.1 illustrates a quark \lq\lq disconnected diagram" 
in lattice QCD. 
These are present in almost all hadron structure 
calculations and represent
one of the greatest calculational challenges in the 
field of lattice QCD. 
Experimental results from Thomas Jefferson National Laboratory
are now closing on on these small effects in the form
of the strange quark contribution to the nucleon 
electromagnetic form factors.

The only previous published lattice results on electric
and magnetic disconnected form factors was in a 1998 paper by Dong, Liu
and Williams\cite{Liu0}. They found that after chiral
extrapolation the disconnected parts gave
\begin{eqnarray*}
G_{M}^{s}(0)=-0.36\pm 0.20,\quad\quad \\
\sqrt{|<r^{2}_{s}>^{n}_{E}|}\approx 0.25
\longleftrightarrow 0.4\, {\rm
fm}.
\end{eqnarray*}
The last number is from monopole fits to the electric
disconnected data. (Measured {\it total} is 
$\sqrt{|<r^{2}>^{n}_{E}|}
=0.35\pm 0.03\, {\rm fm}$).

HAPPEX\cite{HAPPEX} gives 
\begin{eqnarray*}
(G_{E}^{s}+0.392G_{M}^{s})/(G_{M}/\mu_{p})=0.091
\pm 0.54\pm 0.039,
\end{eqnarray*}
at $q^{2}=0.477\,{\rm (Gev/c)^{2}}$. Similarly there 
is a less accurate 
number from SAMPLE\cite{SAMPLE} at $q^{2}=0.1 {\rm (Gev/c)^{2}}$:
\begin{eqnarray*}
G_{M}^{s}(.1 {\rm (Gev/c)^{2}})= 0.61\pm 0.17\pm 0.21.
\end{eqnarray*}
Thus, experimentally there are no solid, separate numbers yet for
disconnected $G^{s}_{M}(q^{2})$ or $G^{s}_{E}(q^{2})$.

Theoretically, results for these quantities from 
different models
are in poor agreement\cite{HAPPEX}.
Preliminary results for disconnected electric and 
magnetic form factors will be
presented here for the case where the disconnected 
quark has the same mass
as the valence quarks. In order to extract the 
strange quark contribution
a chiral extrapolation of the different quark 
mass case must still be done.

\begin{figure}
\begin{center}
\epsfbox{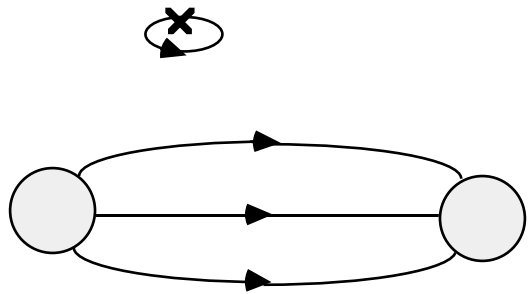}
\vskip .5cm
\caption{A generic nucleon disconnected diagram. 
The time axis is horizontal and 
the shaded circles represent nucleon sources
or sinks.\label{figure1}}
\end{center}
\vskip-1.cm
\end{figure}

\section{Noise and Subtraction Methods}

The basic method used to extract the signal
is $Z2$ noise\cite{Thron} with perturbative noise
subtraction. Consider ${\tilde Q}$ such that
\begin{equation}
<Tr\{{\tilde Q}X \}>=0,
\end{equation}
where $X$ is the noise matrix.
One then has of course for general $Q$
\begin{equation}
<Tr\{(Q-{\tilde Q})X \}>=<Tr\{ QX \}>.
\end{equation}
However, the variance will not be the same,
\begin{equation}
V[Tr\{(Q-{\tilde Q})X \}]\ne V[Tr\{Q X \}].
\end{equation}

The natural choice for ${\tilde Q}$, which must
mimick the off-diagonal components of $Q$ to reduce
the variance, is the {\it perturbative} expansion of the 
quark matrix \cite{Liu1,ww1,michael}. 
This is given by
\begin{eqnarray}
M^{-1}_{p} =  I  + \kappa P  + \kappa^{2} P^{2} +
\kappa^{3} P^{3} + \cdots .
\end{eqnarray}
where for Wilson fermions ($I,J=\{x,a,\alpha\}$)
\begin{eqnarray}
P_{IJ}=\sum_{\mu}[(1+\gamma_{\mu})U_{\mu}(x)
\delta_{x,y-a_{\mu}}+  \nonumber \\
(1-\gamma_{\mu})U_{\mu}^{\dagger}(x-a_{\mu})
\delta_{x,y+a_{\mu}}].
\end{eqnarray}

For a given operator,
${\cal O}$, the matrix ${\cal O}M^{-1}_{p}$ 
encountered in $<{\bar \psi}{\cal
O}\psi>\rightarrow -Tr({\cal O}M^{-1}_{p})$ is 
not traceless. In other words, one
must re-add the perturbative part, subtracted 
earlier, to get the full, unbiased
answer.

Local operators require perturbative
corrections starting at 4th order in $\kappa$
(odd powers need no correction) and point-split 
ones require corrections starting at 3rd order (even
powers need no correction).
This procedure has been implemented exactly 
to 4th order (a single exact subtraction) for the point-split 
conserved vector current on the lattice for
the lowest 4 nonzero lattice momentums.

\section{Analysis Method}

A new analysis method was used for the
disconnected data analysis. At nonzero momentum, the basic 
quantity to consider is a ratio
of nucleon three point and two point funxtions\cite{Liu0},
\begin{equation}
R^{(e,m)}(t,t',\vec{q}) \equiv \frac{ G^{(3e,m)}
(t',t,\vec{q}) }{G^{(2)}(t,0)}
\frac{G^{(2)}(t',0)}{G^{(2)}(t',\vec{q})},
\end{equation}
where $G^{(3e,m)}$ is the appropriate electric or
magnetic three point function and $G^{(2)}(t,\vec{q})$
is the nucleon two point function extending from the time origin
to $t$ with momentum $\vec{q}$. The three point function
comes from the correlation of the two point function and 
the loop data. Calling this Fourier transformed
loop data $L^{(e,m)}(\vec{q},t)=
\sum_{\vec{x}}e^{-i\vec{q}\cdot \vec{x}}J(\vec{x},t)$,
where $J(\vec{x},t)$ is the appropriate component of 
the lattice self-contracted current, then generically
\begin{eqnarray}
G^{(3e,m)}(\vec{q},t',t) =  
<G^{(2e,m)}(t,0)L^{(e,m)}(t',\vec{q})> \nonumber \\
- <G^{(2e,m)}(t,0)><L^{(e,m)}(t',\vec{q})>,
\end{eqnarray}
where the average is over configurations. 
The signal is always in
the imaginary part of the lattice electromagnetic
current $J(\vec{x},t)$, which simplifies 
the analysis and improves
error bars\cite{ww2,Liu2}. 

The actual measurement used here is,
\begin{eqnarray}
M^{(e,m)}(t-\frac{1}{2},\vec{q})\equiv \quad\quad\quad\quad\quad
\quad\quad\quad\nonumber \\
\quad\sum_{t'=1}^{t+1}(R^{(e,m)}(t,t',\vec{q})
-R^{(e,m)}(t-1,t',\vec{q})).\label{eqn8}
\end{eqnarray}
One can show for the appropriate components of 
$G^{(3e,m)}(\vec{q},t',t)$ and $L^{(e,m)}(t',\vec{q})$ that
$M^{(e,m)}(t-\frac{1}{2},\vec{q})\rightarrow G_{E}(\vec{q}^{\,2}),
\quad\frac{q}{E+m}G_{M}(\vec{q}^{\,2})$ 
for $t\gg1$ in the electric or magnetic cases, respectively. 
The jackknife technique was used 
to define correlated error bars and the 
time range $t=10-12$, where the nucleon propagator 
becomes single exponential, 
was fit for all $\vec{q}$. The measurement 
in Eq.(\ref{eqn8}) is 
represented graphically in Fig.6.

The use of the same time-summed loop background in
both terms in (8) leads to significantly smaller
error bars. Eq.(\ref{eqn8}) 
implements the SESAM \lq\lq plateau" 
strategy\cite{SESAM} by requiring the 
quark loop background to 
lengthen in time along with the nucleon
propagators. Linear $t$ fits of the 
individual terms in (8) nicely avoid systematic 
errors coming from \lq\lq off-shell" or contact terms
($t'\sim 0$ or $t'\sim t$, 
constants for $t\gg 1$\,\cite{Gupta})
since the loop time edges do not change their
relation to the nucleon propagator time edges.
However, the statistical errors are substantial. 
In contrast, the two terms in (8) could in principle
possess different off-shell contributions because 
the $t'$ sum in the second term
includes additional correlations with 
loop data two time steps 
beyond the nucleon sink (see Fig.6). It has
been explicitly verified that no discernable 
numerical correlations exist in such data, and so
can be safely included.

\begin{figure}
\begin{center}
\epsfbox{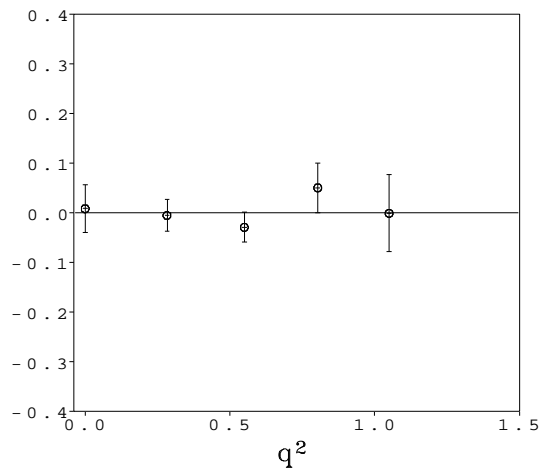}
\vskip-.1cm
\caption{Disconnected nucleon electric form factor
at $\kappa=0.152$ on the $20^{3}\times 32$ lattice as 
a function of $q^{2}$ in $(GeV/c)^{2}$.\label{figure2}}
\end{center}
\vskip-.5cm
\end{figure}

\begin{figure}
\begin{center}
\epsfbox{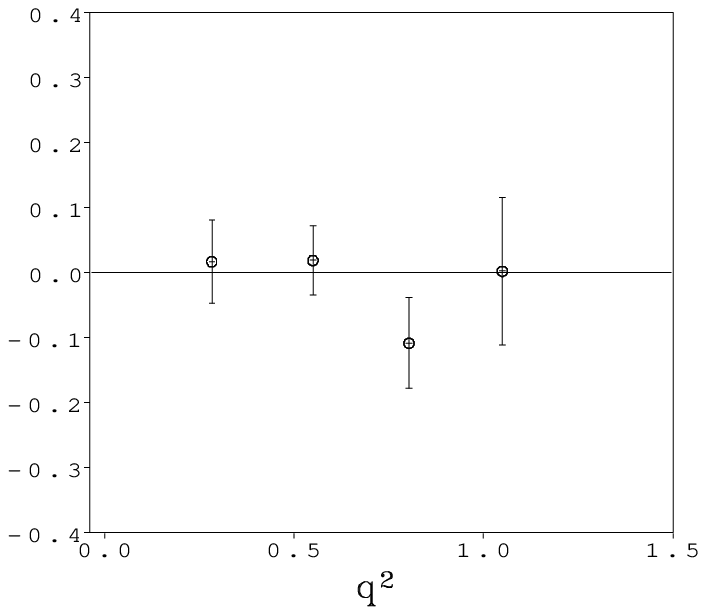}
\vskip-.1cm
\caption{Same as Fig.2 but for
the disconnected nucleon magnetic form factor.
\label{figure3}}
\end{center}
\vskip-.5cm
\end{figure}

\begin{figure}
\begin{center}
\epsfbox{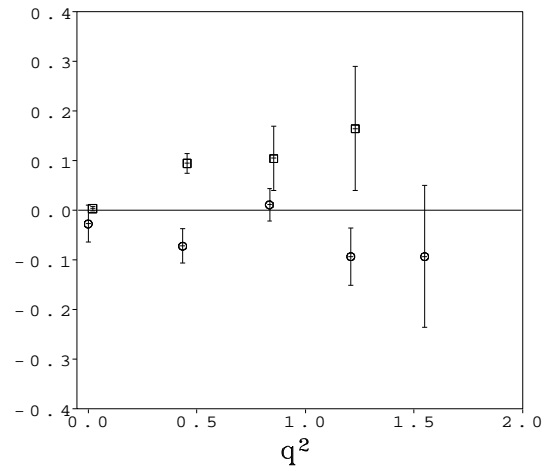}
\vskip-.1cm
\caption{Disconnected nucleon electric form factor
at $\kappa=0.152$ on the $16^{3}\times
24$ lattice as a function of $q^{2}$ in $(GeV/c)^{2}$.
Squares represent previous results from 
Ref.\cite{Liu0}, circles are
present results. Data here and in Fig.\ 5 
separated in $q^{2}$ 
for better visibility.\label{figure4}}
\end{center}
\vskip-.5cm
\end{figure}

\begin{figure}
\begin{center}
\epsfbox{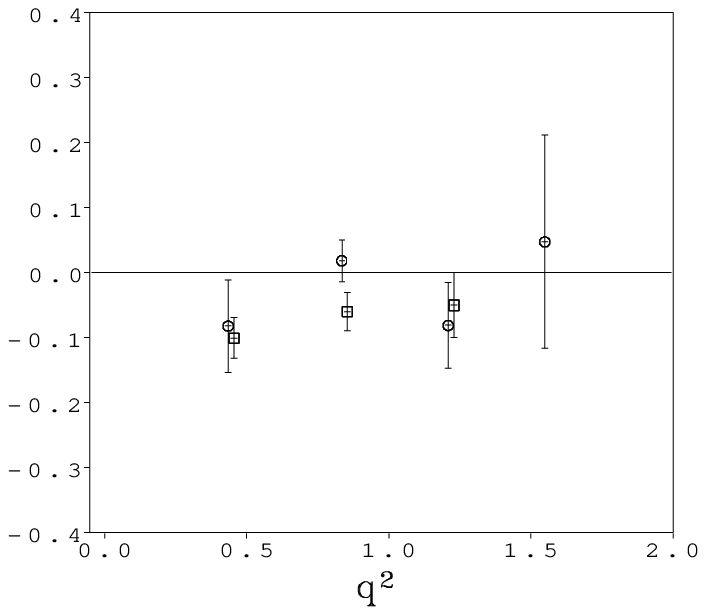}
\vskip-.1cm
\caption{Same as Fig.4 but for the
disconnected nucleon magnetic form factor.
\label{figure5}}
\end{center}
\vskip-.5cm
\end{figure}

\begin{figure}
\begin{center}
\epsfbox{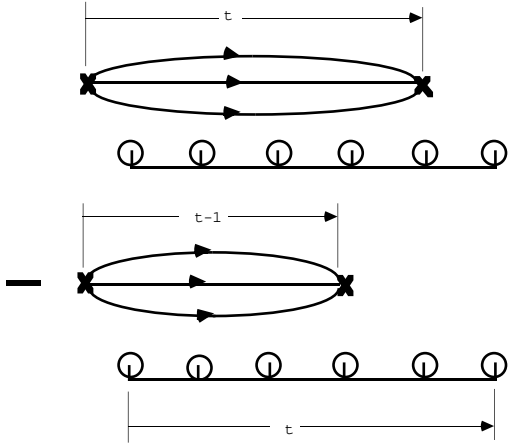}
\vskip-.1cm
\caption{Illustrating the two terms in the disconnected
amplitude measurement, Eq.(\ref{eqn8}). 
The time-summed quark loop data, 
symbolized by the underneath circles,
is correlated with the appropriate nucleon amplitude
and multiplied by a ratio of two point functions.\label{figure6}}
\end{center}
\vskip-.6cm
\end{figure}

\section{Results and Conclusions}

Two lattices sizes were used for comparison 
purposes: $20^{3}\times 32$ and 
$16^{3}\times 24$ at $\beta=6.0$. 
Results at $\kappa=0.152$ on both lattices will be 
examined here (both use 100 configurations).
30 noises on the $20^{3}\times 32$ lattice were used and
60 noises on the $16^{3}\times 24$ lattices. 
As has been shown previously\cite{ww1} at fourth
order subtraction for the electromagnetic current, one gains a 
factor of about 10 in the number of $Z2$ noises at essentially
zero overhead. Thus these results are equivalent to about 
300 unsubtracted noises on the large lattices and about 600
unsubtracted noises on the small lattices.

Results are presented at $\kappa=0.152$ for the $20^{3}
\times 32$ lattices in Figs.2 and 3 above. At this 
$\kappa$ value there is no evidence for a signal
in either the electric or magnetic sectors.
In order to compare more directly with the results
of Ref.\cite{Liu0}, done on 100 $16^{3}\times 24$ lattices
with 300 unsubtracted noises,
the calculation was repeated on identically
sized lattices. Again, there is no evidence of a signal 
at nonzero momentum in the electric or magnetic sectors
in Figs.4 and 5. This finding is in disagreement with 
the Kentucky results\cite{Liu0,Liu2}. The most 
serious disagreement is the electric sector, 
as can be seen in Fig.4.

The main message of these calculations to 
this point is that these
correlations are extremely small QCD effects, both in 
the electric and magnetic sectors. It will be crucial 
to find better methods which will extract the signals
in a more efficient way. Isolating these signals 
and making a definitive
prediction will be one of the greatest challenges
facing hadronic lattice QCD in the years to come.

\section{Acknowledgements}

This work is supported in part by NSF Grant Nos.\ 9722073
and 0070836 and NCSA and utilized the SGI
Origin 2000 System at the University of Illinois. 
The author thanks S.\ J.\ Dong and K.\ F.\ Liu
for their data in Figs.4 and 5.

\end{document}